\documentstyle[12pt,twoside,fleqn,espcrc1]{article}

\newcommand{\AmS}{{\protect\the\textfont2
  A\kern-.1667em\lower.5ex\hbox{M}\kern-.125emS}}

\title{
{\Large 
Parton Cascade Description of Heavy-Ion Collisions at CERN ?}
}

\author{K. Geiger\address{Physics Department\\
	Brookhaven National Laboratory\\
	Upton, N.Y. 11973, U.S.A.}%
}

\input{epsf}

\begin{document}
\maketitle

{\small
\begin{abstract}
There seems to be a general consensus now that a first glimpse of a
QGP-like effect has become visible in the beautiful NA50 data on
$J/\psi$ production and the `anomalous supression' phenomenon.
On the other hand, it is still widely believed that the
dynamics of heavy-ion collisions at CERN SPS energy is
predominantly governed by soft, non-perturbative physics. This is ironic:
after all, it is unlikely that a QGP could be formed if the underlying
dynamics were essentyially soft, rather than that
it requires intense quark-gluon production with sufficient cascade-like
reinteractions which drive the matter to large density and equilibrium.
Therefore, I advocate in this contribution 
that for ultra-relativistic nucleus-nucleus collisions
a description based on
the pQCD interactions and cascade evolution of involved partons
can and should be used, owing to the claim that
short-range parton interactions play
an important role at  sufficiently high beam energies, including
CERN energy  $\sqrt{s} \simeq 20$ A GeV.
Here mini-jet production which liberates of quarks and gluons cannot 
be considered as an isolated rare phenomenon, but can occur
quite copiously and may lead to complex multiple cascade-type processes.
\end{abstract}
}

\section{INTRODUCTION}

A QCD-based space-time model that allows to simulate
nucleus-nucleus collisions (among other particle collisions),
is now available
\footnote{
The program VNI (pronounced {\it Vinnie}) is available
from http://rhic.phys.columbia.edu/rhic/vni.
}
as a computer simulation program called VNI \cite{ms44}.
It is a pQCD parton cascade description \cite{pcm,msrep},
supplemented by a phenomenological
hadronization model \cite{EG},
with dynamically changing proportions of partons and hadrons, in
which the evolution of a nuclear collision is traced
from the first instant of overlap, via QCD parton-cascade development
at the early stage,
parton conversion into pre-hadronic color-singlet clusters and hadron
production through the decays of the clusters,
as well as the fragmentation of the beam remnants at late times.
(I will in the following refer to the parton-cascade/cluster-hadronization
 model as PCM, and to its Monte Carlo implementation as VNI.)
 The PCM description has been
 used \cite{msrep} to provide very useful insight into the dynamics of
  the evolution of the matter at
  energies likely to be reached at BNL RHIC and CERN LHC. The experimental
  data for these will, however, come only in the next millennium.

  Recently, D. Srivastava and myself have taken steps
  to investigate collisions at the CERN SPS and compare the
  model calculations with the increasing body of experimental data,
  thereby addressing the the question:
  {\it Can one really use the PCM picture already at the SPS energies} 
  ($\sqrt{s}\approx$ 17--20 GeV/A)?
  Recall that the parameters of the models were fixed by the
  experimental data for
  $pp$ ($p\overline{p}$) cross-sections over $\sqrt{s}$ = 10--1800 GeV, and
  $e^+e^-$ annihilation \cite{msrep}. The nucleon-nucleon energy reached at SPS is
  well within this range.
  Indeed, as shown in \cite{dks1,dks2}, this approach does remarkably well in comparison
  to the gross particle production properties observed at CERN SPS.
  This came as a surprise, since
  no attempt was made to fine-tune the model to the data.

\section{COMPARISON WITH CERN SPS DATA}

In Figs. 1 and 2, I have plotted  examples of the results of \cite{dks1,dks2},
for the relatively `light' system $S +S$ and for the most `heavy' system $Pb+Pb$.
The first plot shows the rapidity distribution of negative hadrons in 
central collisions of sulfur nuclei, whereas the second plot shows the transverse
mass spectra of negative hadrons in central lead-lead colisions.
In comparison with the corresponding experimental data,
it is evident that a decent description is obtained, and this
without any adjustments of parameters.
The  analysis of \cite{dks1} concluded:
\begin{description}
\item{(i)}
The simulations of heavy-ion collsions give
a  good overall description of the CERN SPS data on rapidity and
transverse energy distributions,
the multiplicity distributions, as well as a reasonable
description of the transverse momentum distribution of hadrons.
This success may be taken as a `post-hum' justification
of the applicability of a pQCD parton decription even at CERN SPS energies.
\item{(ii)}
Contrary to wide-spread belief,
at CERN SPS energy
the `hard' (perturbative)  parton production and parton cascading
is an important element for particle production
at central rapidities - at least in $Pb+Pb$ collisions. In effect
it provides almost 50 $\%$ of final particles around mid-rapidity.
`Soft' (non-perturbative)  parton interactions
on the other hand are found to be insignificant in their
effect on final-state particle distibutions.
\end{description}
\vspace{-0.5cm}

\begin{figure}[htb]
\begin{minipage}[t]{80mm}
\epsfxsize=200pt
\rightline{ \epsfbox{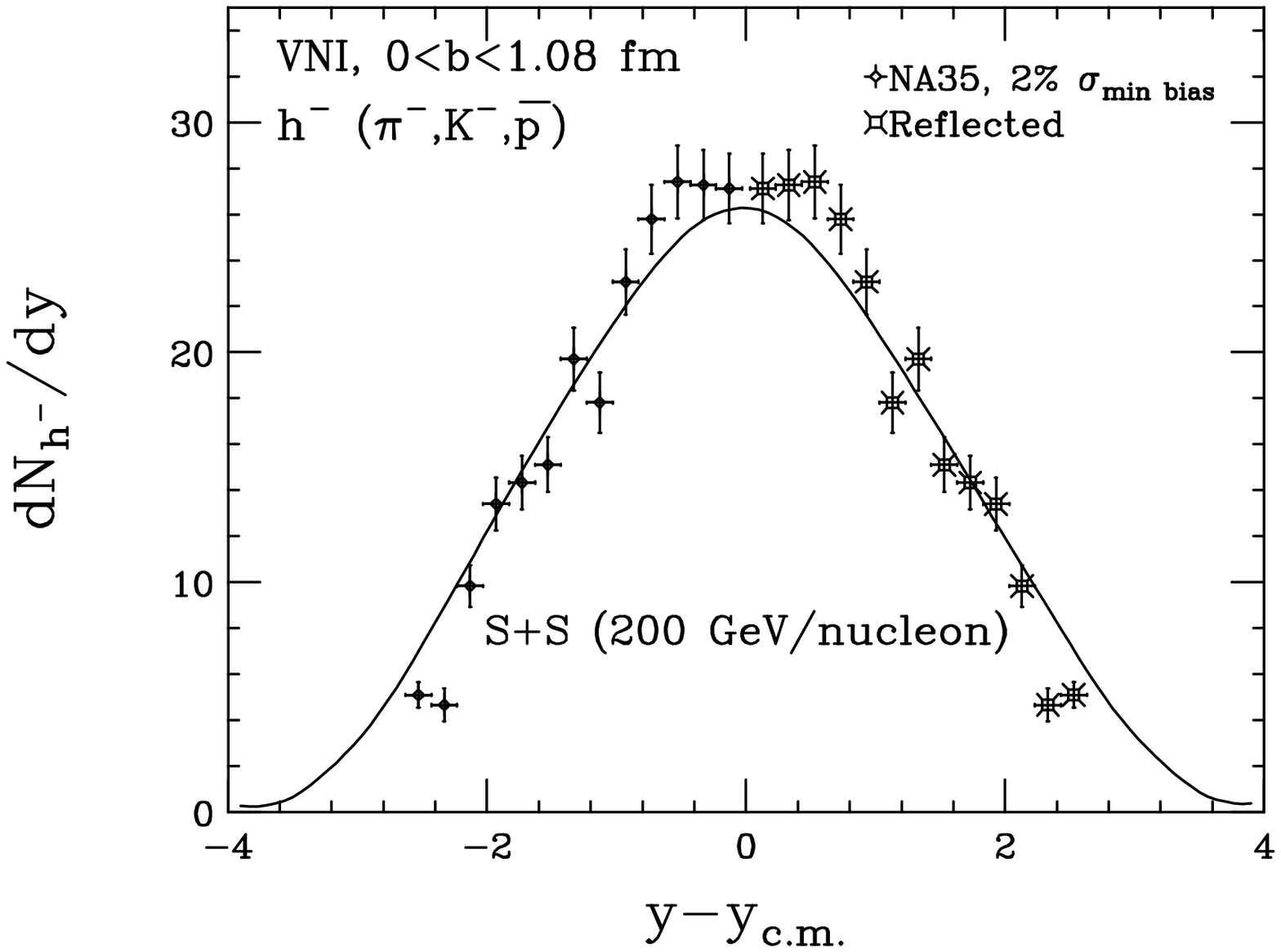} $\;\;\;\;\;\;\;\;\;$  }
\vspace{-0.8cm}
\caption{ The rapidity distribution of negative hadrons for most
central collisions of sulfur nuclei at CERN SPS \protect\cite{dks2}.}
\end{minipage}
\hspace{\fill}
\begin{minipage}[t]{75mm}
\epsfxsize=200pt
\centerline{ \epsfbox{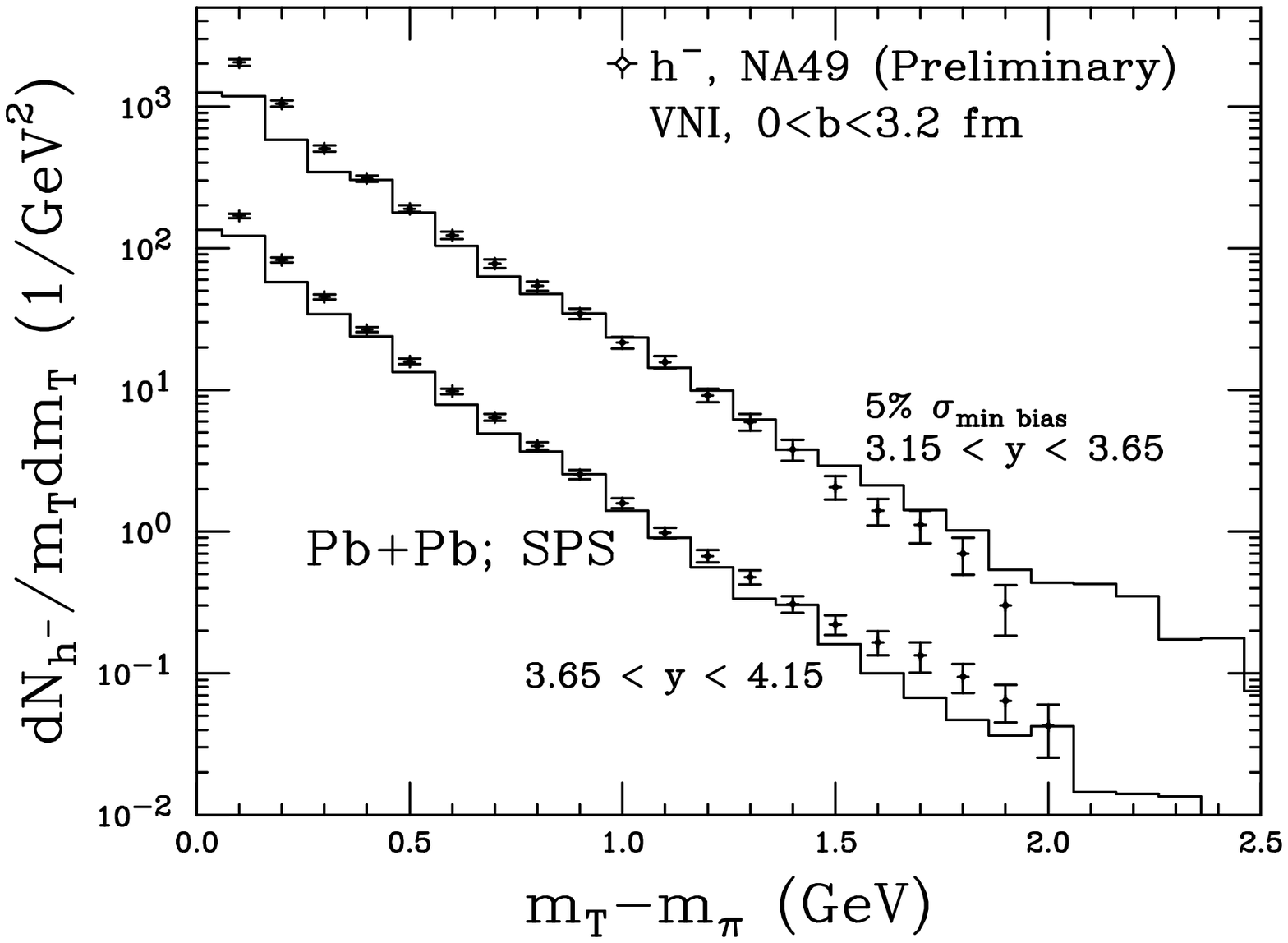} $\;\;\;$ }
\vspace{-0.8cm}
\caption{The transverse mass spectra of negative hadrons
($\pi^-$, $K^-$, and $\overline{p}$)
successively scaled down by an order of magnitude. The predictions are
normalized to the data ar $m_T-m_\pi$= 0.9 GeV \protect\cite{dks1}.}
\end{minipage}
\end{figure}
\vspace{-0.7cm}

\section{IMPLICATIONS FOR {\boldmath $J/\psi$} SUPPRESSION}

Let me now turn to the problem of the observed strong suppression of
$J/\psi$ production in Pb + Pb collisions, following a 
recent work \cite{ms49} by B. M\"uller and myself.  In particular, I would like to
discuss briefly the $A$-dependence of the absorption of the
$J/\psi$ by comoving produced hadronic matter. 
The key element in the following arguments is the fact that,
for most nuclear collision systems
at the relatively low center-of-mass energy of the CERN-SPS
experiments, perturbative mini-jet production is largely
due to quark-quark scattering, because the typical value of
Bjorken-$x$ probed is $\langle x \rangle
\, \lower3pt\hbox{$\buildrel >\over\sim$}\,0.1$, where the gluon density
is small.  This implies that shadowing effects are negligible
as explained in \cite{ms49}
and the critical momentum $p_{crit}$ where these effects turn on,
lies below the perturbatively accessible range of momenta
$p_\perp \ge 1$ GeV at the SPS energy, and may just begin to reach into
it for the Pb + Pb system.
If this is correct, then {\it all} mini-jet production involving momenta
$p_\perp \ge 1$ GeV lies safely above $p_{\rm crit}$.
As immediate consequence, the density of
comovers which can effectively interact and absorb the $J/\psi$ at the
SPS grows like $A^{2/3}$, or $(A_1A_2)^{1/3}$ in asymmetric
collisions.  This is a {\it much stronger $A$-dependence} than naively
expected and embodied in most comover suppression models \cite{Gavin}.
It also implies a {\it stronger impact parameter or $E_T$-dependence} of
comover suppression than predicted by existing models.

Quantitative support for these speculations comes from recent calculations
of secondary particle production in the parton cascade
model, usisng VNI \cite{ms44}. 
The calculations predict that the energy density $\epsilon$ produced
by scattering partons (mini-jets) at central rapidity grows by a factor
of more than 2 between $S + U$ and $Pb + Pb$, compatible with the scaling
law $\epsilon\sim(A_1A_2)^{1/3}$, but much faster than expected in
the usual Glauber model approach which predicts $\epsilon\sim
(A_1^{1/3}+A_2^{1/3})$, in which case $\epsilon$ increases only by 30 $\%$.
\vspace{-0.7cm}

\begin{figure}[htb]
\begin{minipage}[t]{80mm}
\epsfxsize=250pt
\rightline{ \epsfbox{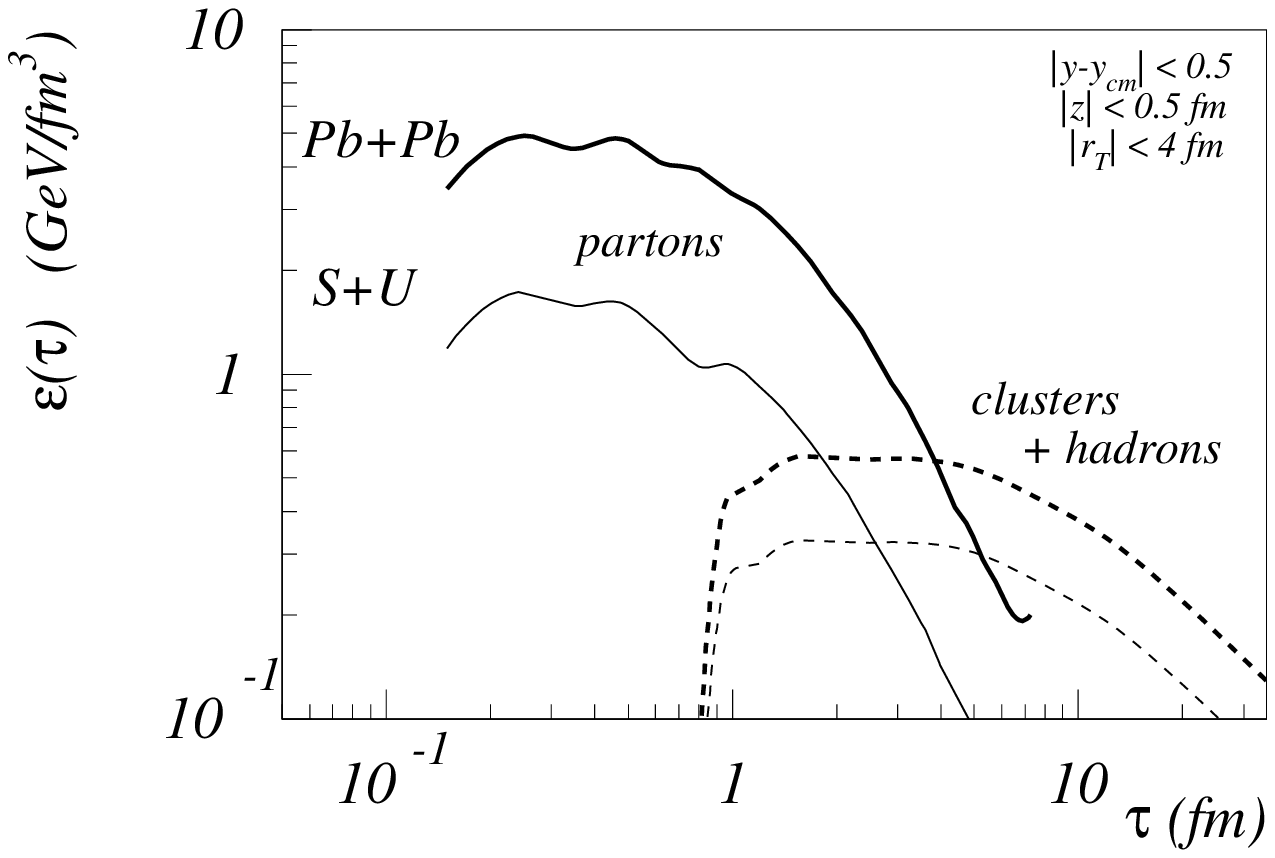}  }
\vspace{-0.9cm}
\caption{ 
Time evolution of the energy density $\epsilon$ of the 
partonic matter in the central slice of the collision systems 
$S + U$ and $Pb + Pb$ at CERN-SPS
beam energies of 200 $A\cdot$GeV and 158 $A\cdot$GeV \protect\cite{ms49}.
}
\end{minipage}
\hspace{\fill}
\begin{minipage}[t]{75mm}
\epsfxsize=250pt
\centerline{ \epsfbox{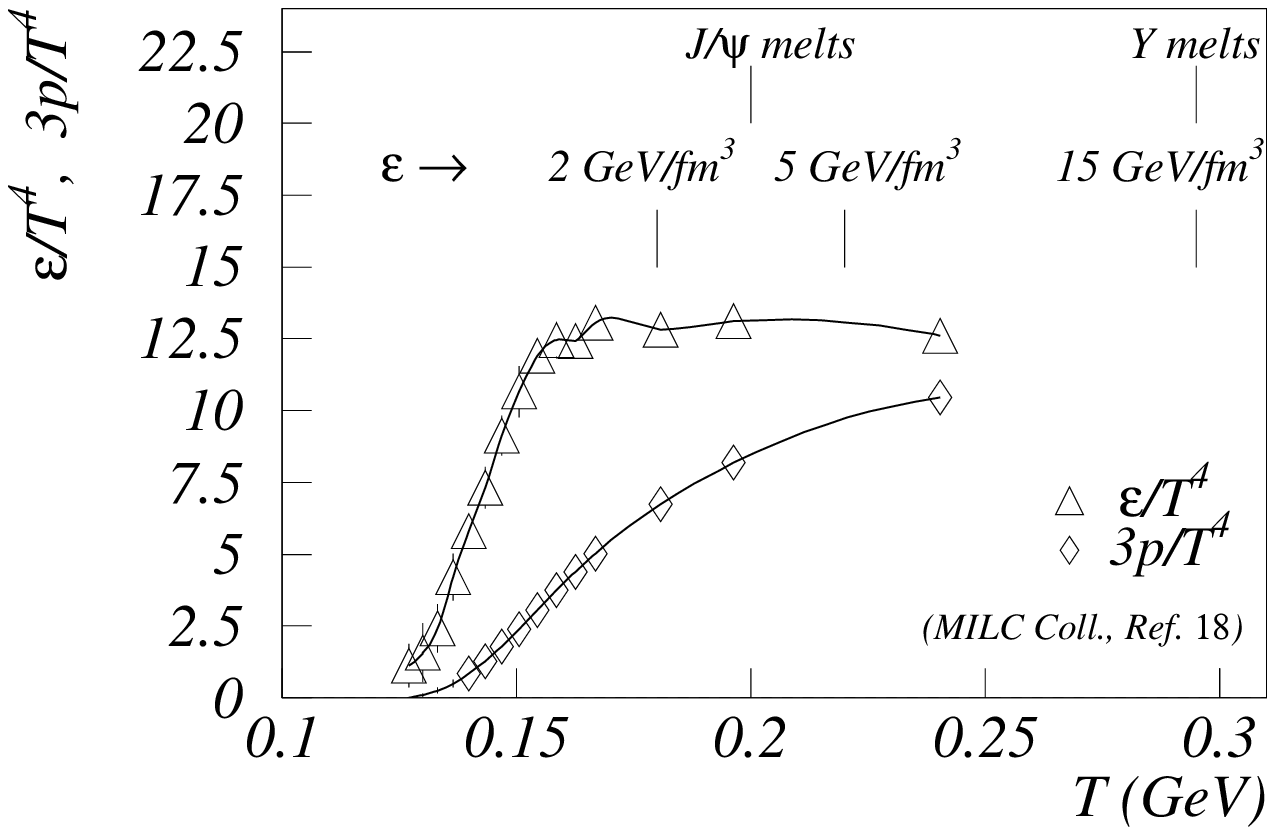} $\;\;\;$ }
\vspace{-0.9cm}
\caption{
Equation of state for two-flavor QCD 
\protect\cite{Blum}, showing energy density
$\epsilon$ and pressure $p$ as a function of temperature
$T$ and coreponding energy densities with
the location of the ''melting'' points of the $J/\psi$ and $\Upsilon$
states \protect\cite{KS91}.
}
\end{minipage}
\end{figure}
\vspace{-0.3cm}
In $Pb + Pb$ collisions the initial partonic energy density $\epsilon$
is predicted to reach 5 GeV/fm$^3$ at times $\tau < 1$ fm/$c$ in the
comoving reference frame (Fig. 3).  For a thermalized gas of free
gluons and three flavors of light quarks this corresponds to an initial
temperature $T_i \approx 230$ MeV, clearly above the critical
temperature $T_c\approx 150$ MeV predicted by lattice-QCD calculations
\cite{Blum} shown in Fig. 4.  On the other hand, the parton density
$\epsilon \approx$ 2 GeV/fm$^3$ predicted for $S + U$ collisions (Fig. 3),
just lies at the upper end of the transition region in the equation of
state from lattice-QCD.

The temperature $T_D$ required for dissociation of the $J/\psi$ bound
state due to color screening is known \cite{KS91} to be higher than
$T_c$, namely $T_D\approx 1.2 \,T_c = 180-200$ MeV.  The parton cascade model
results are therefore compatible with the experimental finding that there
appears to exist no significant comover-induced suppression of $J/\psi$ in
$S + U$ collisions, but a large and strongly impact parameter dependent effect
is observed in Pb + Pb collisions.

It must be stressed that the model results do not contradict
the experimental data collected at the SPS, as one might suspect,
since it is usually claimed that those are fully consistent with
the Glauber picture, in particular with the transverse energy production
(Fig. 5) and the linear $E_\perp-E_{veto}$ relation (Fig. 6), both
wich are in good agreement with the corresponding measured data \cite{NA49}.
\vspace{-0.9cm}

\begin{figure}[htb]
\begin{minipage}[t]{80mm}
\epsfxsize=150pt
\rightline{ \epsfbox{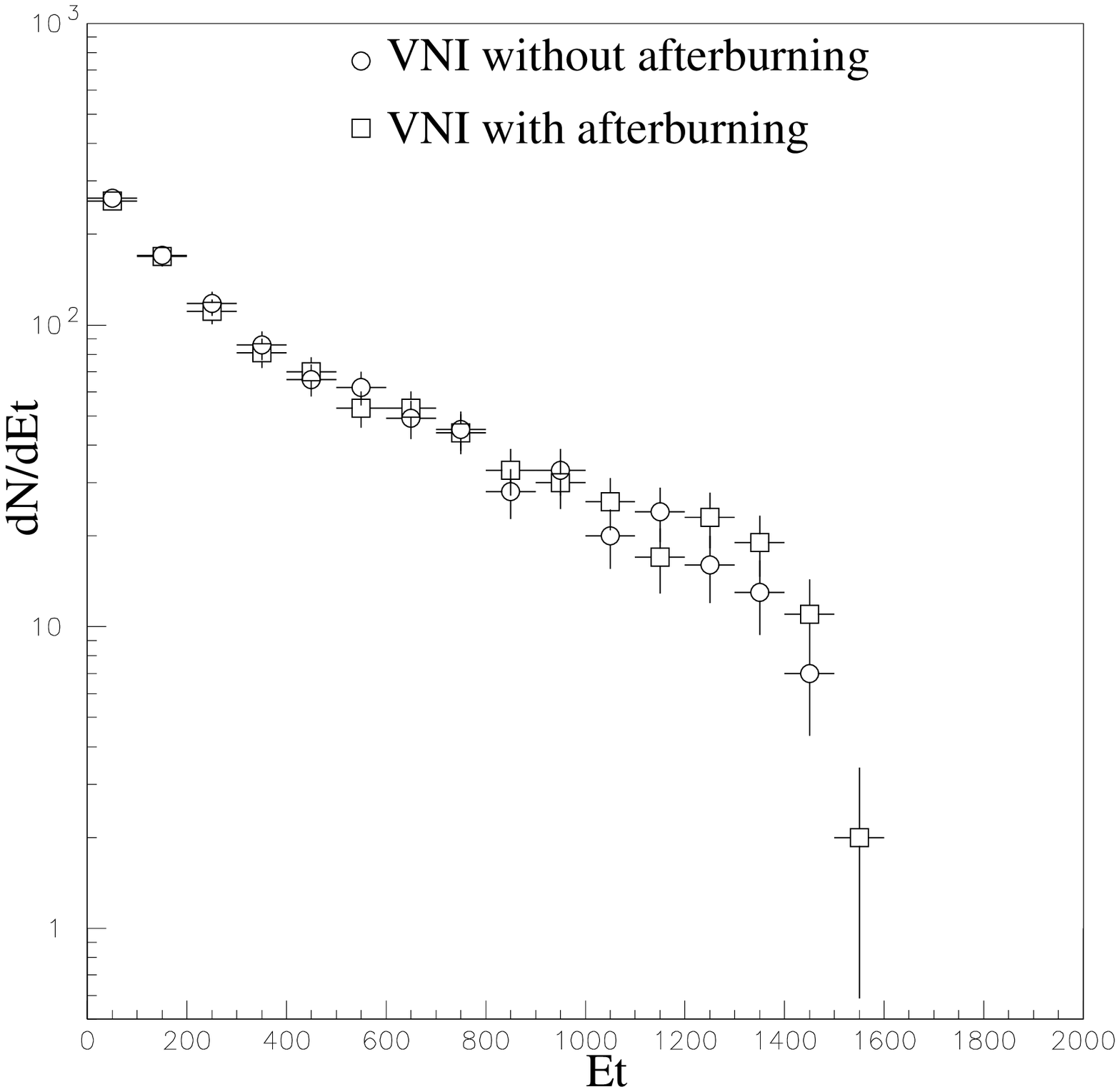} $\;\;\;\;\;\;\;\;\;$  }
\vspace{-1.2cm}
\caption{ Distribution of transverse energy production in minimum
bias collsions of $Pb+Pb$ at CERN SPS, as calculated with VNI \protect{\cite{ms44}}.}
\end{minipage}
\hspace{\fill}
\begin{minipage}[t]{75mm}
\epsfxsize=200pt
\centerline{ \epsfbox{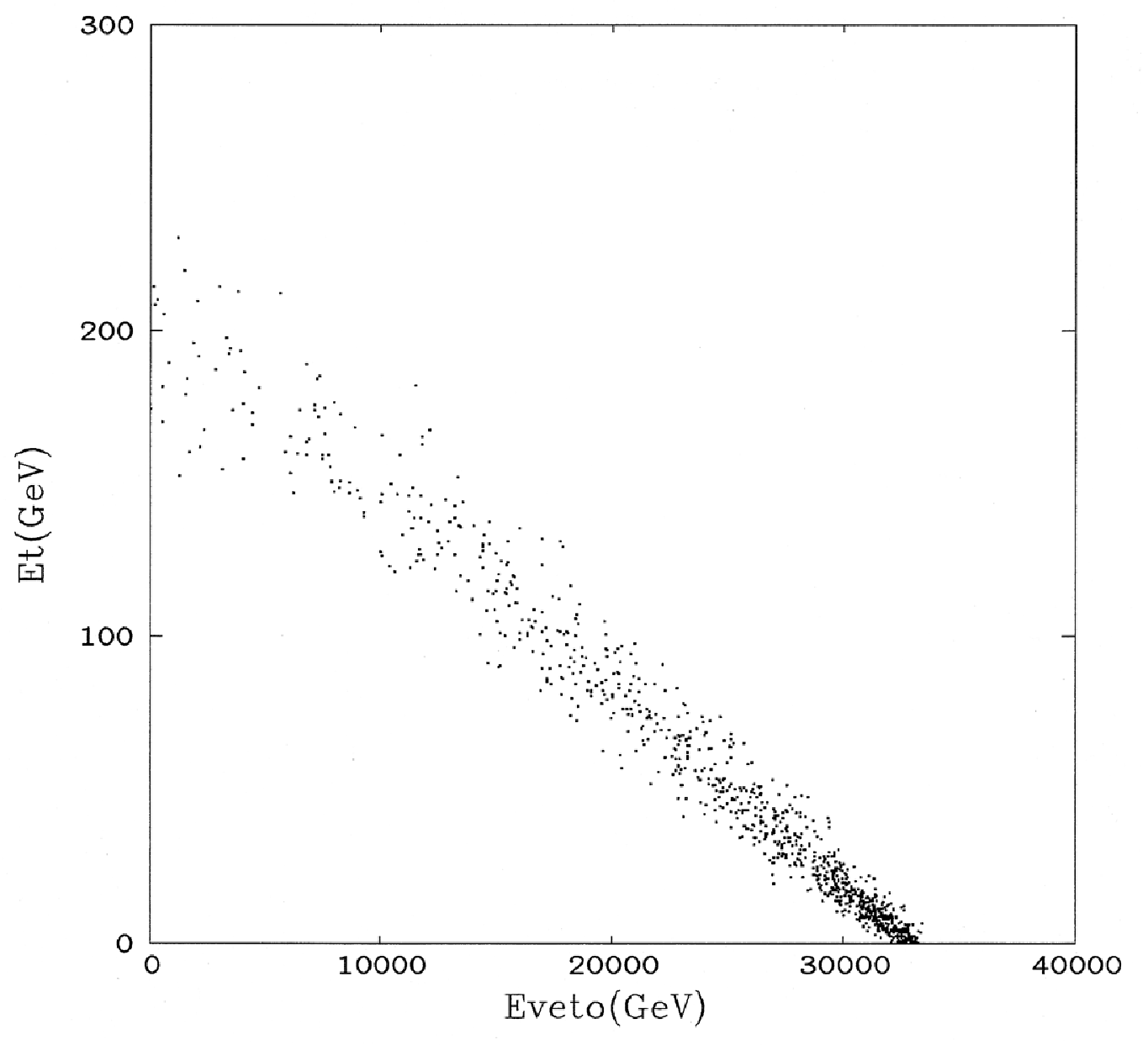} $\;\;\;$ }
\vspace{-0.7cm}
\caption{For $1.1 \le\eta\le 2.3$, the correlation between $E_{veto}$, 
the energy deposited in the forward beam calorimeter, and $E_\perp$, 
the energy harnessed in transverse direction, from VNI.}
\end{minipage}
\end{figure}
\noindent{\bf ACKNOWLEDGEMENT}

\noindent
This work was supported in part by grant
DE-AC02-76H00016 from the U.S.~Department of Energy.

\end{document}